\documentclass[reprint,amsmath,amssymb,aps,prl]{revtex4-1}
\usepackage{graphicx}
\usepackage{psfrag}
\usepackage{dcolumn}
\usepackage{bm}
\begin{document}

\title{Ordering of the Heisenberg spin glass in four dimensions}

\author{Hikaru Kawamura}
\email{kawamura@ess.sci.osaka-u.ac.jp}
%\altaffiliation{Present Address:
%Department of Earth and Space Science, Faculty of Science,
%Osaka University, Toyonaka, Osaka 560-0043, Japan}
%\affiliation{Department of Earth and Space Science, Faculty of Science,
%Osaka University, Toyonaka, Osaka 560-0043, Japan}

\author{Shinichirou Nishikawa}
%\altaffiliation{Present Address:
%Department of Earth and Space Science, Faculty of Science,
%Osaka University, Toyonaka, Osaka 560-0043, Japan}

\affiliation{Graduate School of Science, Osaka University,
Toyonaka, Osaka 560-0043, Japan}

\date{\today}

\begin{abstract}
 Ordering of the Heisenberg spin glass in four dimensions (4D) with the nearest-neighbor Gaussian coupling is investigated by equilibrium Monte Carlo simulations, with particular attention to its spin and chiral orderings. It is found that the spin and the chirality order simultaneously with a common correlation-length exponent $\nu_{CG}=\nu_{SG}\simeq 1.0$, {\it i.e.\/}, the absence of the spin-chirality decoupling in 4D. Yet, the spin-glass ordered state exhibits a nontrivial phase-space structure associated with a continuous one-step-like replica-symmetry breaking, different in nature from that of the Ising spin glass or of the mean-field spin glass. Comparison is made with the ordering of the Heisenberg spin glass in 3D, and with that of the 1D Heisenberg spin glass with a long-range power-law interaction. It is argued that the 4D might be close to the marginal dimension separating the spin-chirality decoupling/coupling regimes.
\end{abstract}
\maketitle

%%%%%%%%%%%%%%%%%
% INTRODUCTION %
%%%%%%%%%%%%%%%%%

\section{I. Introduction}

 The Heisenberg spin-glass (SG) model, or the Edwards-Anderson model with the isotropic Heisenberg exchange interaction \cite{EA}, has been considered as the standard model of many real SG materials \cite{review}. In realistic three spatial dimensions (3D), earlier studies in the 80's suggested that the isotropic Heisenberg SG did not exhibit an equilibrium SG transition at any finite temperature in apparent contrast to experiments \cite{Banavar82,McMillan85,Olive86,Matsubara91,Yoshino93}. Then, a proposal was made  in 1992 that the model might exhibit a finite-temperature transition in the chiral sector, with the standard SG order occurring at a temperature $T_{SG}$ lower than the chiral-glass (CG) ordering temperature $T_{CG}$, {\it i.e.\/}, $T_{SG} < T_{CG}$ \cite{Kawamura92}. The occurrence of such separate spin and chirality transitions is now called ``spin-chirality decoupling'' \cite{Kawamura10}. Chirality is a multispin variable representing the sense or the handedness of local noncoplanar spin structures induced by spin frustration. A possible counter-view to such a picture might be that the spin and the chirality order at a common finite temperature \cite{Matsubara,LeeYoung03,Campos06,LeeYoung07,Fernandez09}. Although there is no complete consensus, recent simulations point to the occurrence of the spin-chirality decoupling in 3D \cite{Matsumoto02,HukuKawa05,Campbell07,VietKawamura09a,VietKawamura09b}. For example, Ref.\cite{VietKawamura09b} reported that $T_{SG}$ was lower than $T_{CG}$ by about $10\sim 15$\%. 

 To get further insight into the issue, it might be useful to extend the space dimension from original $d=3$ to general $d$-dimensions. In $d=1$, the  Heisenberg SG with a short-range (SR) interaction exhibits only a $T=0$ transition. In $d=2$, recent calculations suggested that the vector SG model, either the three-component Heisenberg SG \cite{KawaYone03} or the two-component {\it XY\/} SG \cite{Weigel08}, exhibited a $T=0$ transition but with the spin-chirality decoupling, {\it i.e.\/}, the CG correlation-length exponent $\nu_{CG}$ was greater than the SG correlation-length exponent $\nu_{SG}$, meaning that this $T=0$ transition was characterized by two distinct diverging length scales, each associated with the chirality and with the spin. In the opposite limit of $d\rightarrow \infty$, the model is known to reduce to the mean-field (MF) model, {\it i.e.\/}, the Sherrington-Kirkpatrick (SK) model. The Heisenberg SK model is known to exhibit a single finite-temperature transition, with no spin-chirality decoupling. In high but finite $d$, Monte Carlo (MC) study by Imagawa and Kawamura suggested that the spin-chirality decoupling did not occur for $d=5$, whereas the situation in $d=4$ appeared somewhat more marginal \cite{ImaKawa03a}. 

 Another useful way of attacking the issue might be to study the one-dimensional (1D) Heisenberg SG with a long-range (LR) power-law interaction proportional to $1/r^{\sigma}$ ($r$ the spin distance). Indeed, several studies both for the Ising and the Heisenberg SGs suggested that the 1D LR SG model with a power-law exponent $\sigma$ might show the ordering behavior analogous to the $d$-dimensional SG model with a SR interaction \cite{BinderYoung86,Katzgraber03,Katzgraber05,Katzgraber09,Leuzzi99,Leuzzi08,VietKawamura10a, VietKawamura10b,SharmaYoung}. Even a simple empirical formula relating $\sigma$ and $d$, $d=2/(2\sigma-1)$, was proposed \cite{BinderYoung86}, though the relation is only approximate. 

 Recent MC calculation on the 1D LR Heisenberg SG by Viet and Kawamura suggested that the spin-chirality decoupling occurred for $\sigma \geq \sigma_c$, but did not occur for $\sigma \leq \sigma_c$, $\sigma_c$ being estimated numerically to be $\sigma_c\simeq 0.8$ \cite{VietKawamura10a,VietKawamura10b}. If one applies the approximate $d-\sigma$ correspondence formula quoted above \cite{BinderYoung86}, the critical dimension below which the spin-chirality decoupling is expected would be $d_c\simeq 3.3$, suggesting that $d=4$ might lie near the margin of, slightly on the side of the spin-chirality coupling regime. Of course, the above $d-\sigma$ correspondence formula is only approximate, and even whether $d=4$ is greater or smaller than $d_c$ is not clear. Previous simulation on the $4d$ Heisenberg SG, which dealt with the linear size of $L\leq 10$, was not definitive concerning the occurrence of the spin-chirality decoupling in 4D \cite{ImaKawa03a}.

 Under such circumstances, the purpose of the present paper is first to clarify whether the spin and the chirality are decoupled or not in the $4d$ Heisenberg SG by simulating larger systems than the ones studied in Ref.\cite{ImaKawa03a}. Since $d=4$ is expected to be close to the marginal dimension concerning the spin-chirality decoupling, we wish to see what kind of ordering behavior is realized for the spin and the chirality near the marginal dimension.

 The present paper is organized as follows. In \S II, we introduce our model and explain some of the details of our MC simulation. Various physical quantities are defined in \S III. The results of our MC simulations, including the spin and the chiral correlation lengths, the spin and the chiral Binder ratios,  are presented in \S IV. The SG and CG transition temperatures are determined by carefully examining the size dependence of the finite-size data. In \S V, critical properties of the spin and of the chirality are investigated by means of a finite-size scaling analysis. Finally, section \S VI is devoted to summary and discussion.

\section{II. The model and the method}
\label{secModel}

The model we consider is the isotropic classical Heisenberg model on a 4D hyper-cubic lattice, with the nearest-neighbor Gaussian coupling. The Hamiltonian is given by
\begin{equation}
{\cal H}=-\sum_{<ij>}J_{ij}\vec{S}_i\cdot \vec{S}_j\ \ , 
\label{eqn:hamil}
\end{equation}
where $\vec{S}_i=(S_i^x,S_i^y,S_i^z)$ is a three-component unit vector, and $<ij>$ sum is taken over nearest-neighbor pairs on the lattice. The nearest-neighbor coupling $J_{ij}$ is assumed to obey the Gaussian distribution with zero mean and variance $J^2$, which is taken to be unity ($J=1$) in the following. The temperature $T$ is measured in units of $J$.

 We perform equilibrium MC simulations on the model. The lattices are hyper-cubic lattices with $N=L^{4}$ sites with $L=6$, 8, 10, 12, 16 and 20. We impose periodic boundary conditions in all four directions. Sample average is taken over 1300, 1200, 840, 590, 430 and 256 independent bond realizations for $L=6$, 8, 10, 12, 16 and 20, respectively. Error bars of physical quantities are estimated by sample-to-sample statistical fluctuations over the bond realization.

 In order to facilitate efficient thermalization, we combine the heat-bath and the over-relaxation methods with the temperature-exchange technique \cite{HukushimaNemoto}. For each heat-bath sweep we perform 11, 15, 19, 23, 31 and 55 over-relaxation sweeps, while the total number of temperature points in the temperature-exchange process are taken to be 35, 51, 59, 55, 55 and 60 for $L=6$, 8, 10, 12, 16 and 20, respectively. Care is taken to be sure that the system is fully equilibrated. Equilibration is checked by following the procedures of Ref.\cite{VietKawamura09b}.

\section{III. Physical quantities}

In this section, we define various physical quantities measured in our simulations.

For the Heisenberg spin, the local chirality at the $i$-th site and in the $\mu$-th direction $\chi_{i\mu}$ may be defined for three neighboring Heisenberg spins by a scalar
\begin{equation}
\chi_{i\mu}=
\vec{S}_{i+{\hat{e}}_{\mu}}\cdot
(\vec{S}_i\times\vec{S}_{i-{\hat{e}}_{\mu}}),
\end{equation}
where ${\hat{e}}_{\mu}\ (\mu=x,y,z,u)$ denotes a unit vector along the $\mu$-th axis. There are in total $4N$ local chiral variables.

 We define an ``overlap'' for the chirality. We prepare at each temperature two independent systems 1 and 2 described by the same Hamiltonian (1) with the same interaction set. We simulate these two independent systems 1 and 2 in parallel with using different spin initial conditions and different sequences of random numbers.  

The $k$-dependent chiral overlap, $q_\chi(\vec k)$, is defined as an overlap variable between the two replicas 1 and 2 as a scalar
\begin{equation}
q_\chi(\vec k) =
\frac{1}{4N}\sum_{i=1}^N\sum_{\mu=x,y,z,u}
\chi_{i\mu}^{(1)}\chi_{i\mu}^{(2)}e^{i\vec k\cdot \vec r_i},
\end{equation}
where the upper suffixes (1) and (2) denote the two replicas of the system, and $\vec r_i$ is the position vector of the site $i$.

The $k$-dependent spin overlap, $q_{\alpha\beta}(\vec k)$, is defined by a {\it tensor\/} variable between the $\alpha$ and $\beta$ components of the Heisenberg spin,
\begin{equation}
q_{\alpha\beta}(\vec k) = 
\frac{1}{N}\sum_{i=1}^N S_{i\alpha}^{(1)}S_{i\beta}^{(2)}e^{i\vec k\cdot \vec r_i},
\ \ \ (\alpha,\beta=x,y,z).
\end{equation}

In term of the $k$-dependent overlap, the CG and the SG order parameters are defined by the second moment of the overlap at a wavevector $k=0$,
\begin{equation}
q_{CG}^{(2)}=\frac {[\langle | q_{\chi}(\vec 0)|^2 \rangle]} {\overline{\chi}^{4}},
\end{equation}
\begin{equation}
q_{SG}^{(2)} = [\langle q_{\rm s}(\vec 0)^2\rangle]\ ,
\ \ \  
q_{\rm s}(\vec k)^2 = \sum_{\alpha,\beta=x,y,z} \left| q_{\alpha\beta}(\vec k) \right| ^2,
\end{equation}
where $\langle \cdots \rangle$ represents a thermal average and $[\cdots ]$ an average over the bond disorder. The CG order parameter $q_{CG}^{(2)}$ has been normalized here by the mean-square amplitude of the local chirality,
\begin{equation}
\overline{\chi}^{2}=\frac{1}{4N}\sum_{i=1}^N\sum_{\mu=x,y,z,u}[\langle \chi_{i\mu}^2\rangle],
\end{equation}
which remains nonzero only when the spin has a noncoplanar structure locally. The local-chirality amplitude depends on the temperature and the lattice size only weakly.

%The CG and SG susceptibilities are defined by 
%
%\begin{equation}
%\chi_{CG}=3Nq_{CG}^{(2)}\ , \ \ \  \chi_{SG}=Nq_{SG}^{(2)}.
%\end{equation}
%

Finite-size correlation length $\xi_L$ is defined by
\begin{equation}
\xi_L = 
\frac{1}{2\sin(k_\mathrm{m}/2)}
\sqrt{ \frac{ [\langle q(\vec 0)^2 \rangle] }
{[\langle q(\vec{k}_\mathrm{m})^2 \rangle] } -1 },
\end{equation}
for each case of the chirality and the spin, $\xi_{CG}$ and $\xi_{SG}$, where $\vec{k}_{\rm m}=(2\pi/L,0,0,0)$ with $k_{\textrm{m}}=|\vec k_{\textrm{m}}|$. For the CG correlation length $\xi_{CG}$, we consider two distinct definitions depending on the mutual direction between the $\hat{e}_\mu$-vector appearing in the definition of the local chirality (2) and the $\vec{k}_\mathrm{m}$-vector. When $\hat{e}_\mu \parallel \vec{k}_\mathrm{m}$, {\it i.e.\/}, $\mu=x$, we call the corresponding $\xi_{CG}$ the parallel CG correlation length $\xi_{CG}^\parallel$, whereas, when $\hat{e}_\mu \perp \vec{k}_\mathrm{m}$, {\it i.e.\/}, $\mu=y,z,u$, we call the corresponding $\xi_{CG}$ the perpendicular CG correlation length. The perpendicular CG correlation length $\xi_{CG}^\perp$ is actually defined by the mean of three equivalent ones, each defined in the $\mu=y,z,u$ directions.

The CG and the SG Binder ratios are defined by
\begin{equation}
g_{CG}=
\frac{1}{2}
\left(3-\frac{[\langle q_{\chi}(\vec 0)^4\rangle]}
{[\langle q_{\chi}(\vec 0)^2\rangle]^2}\right),
\end{equation}
\begin{equation}
g_{SG} = \frac{1}{2}
\left(11 - 9\frac{[\langle q_{\rm s}(\vec 0)^4\rangle]}
{[\langle q_{\rm s}(\vec 0)^2\rangle]^2}\right).
\label{eqn:gs_def}
\end{equation}
These quantities are normalized so that, in the thermodynamic limit, they vanish in the high-temperature phase and gives unity in the non-degenerate ordered state. In the present Gaussian coupling model, the ground state is expected to be non-degenerate so that both $g_{CG}$ and $g_{SG}$ should be unity at $T=0$.

\section{IV. Monte Carlo results}

In this section, we present the results of our MC simulations on the 4D isotropic Heisenberg SG with the random Gaussian coupling.

 We show in Fig.1 the temperature dependence of the CG and the SG correlation-length ratios, $\xi_{SG}/L$ in (a), $\xi_{CG}^\perp/L$ in (b), and $\xi_{CG}^\parallel/L$ in (c). As can be seen from the figures, both the chiral $\xi_{SG}/L$ and the spin  $\xi_{CG}/L$ curves cross at temperatures which are weakly $L$-dependent. Magnified views of the crossing-temperature range are shown in Figs.2(a)-(c) for the spin, the perpendicular chirality and the parallel chirality, respectively.

\begin{figure}[ht]
\begin{center}
\includegraphics[scale=1.1]{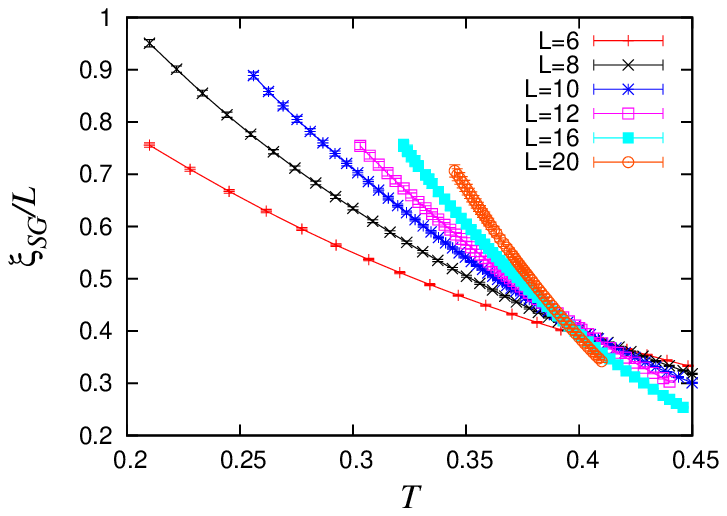}
\includegraphics[scale=1.1]{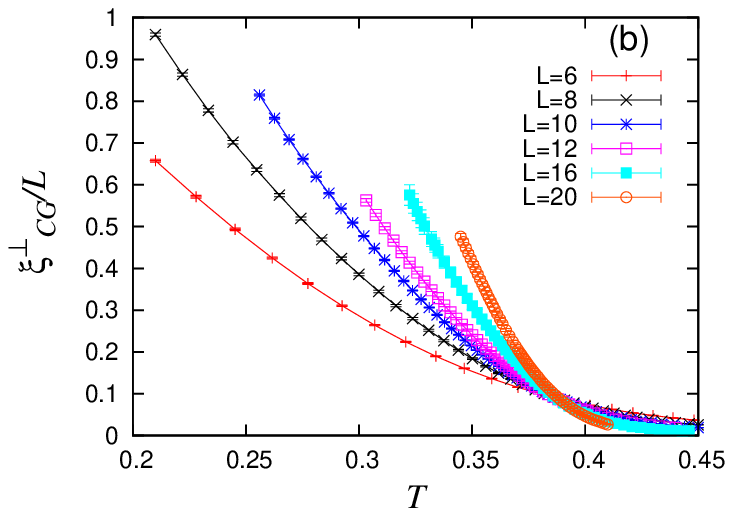}
\includegraphics[scale=1.1]{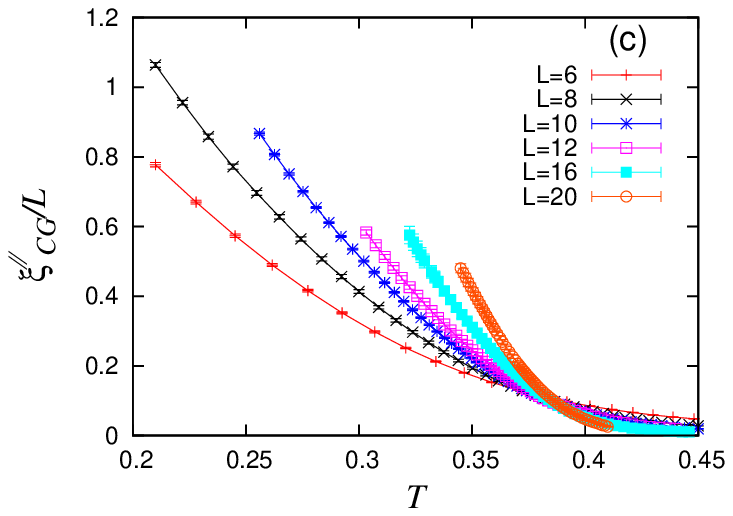}
\end{center}
\caption{
(Color online) The temperature and size dependence of the spin correlation-length ratio (a), of the perpendicular chiral correlation-length ratio (b), and of the parallel chiral correlation-length ratio (c). 
}
\end{figure}
\begin{figure}[ht]
\begin{center}
\includegraphics[scale=1.1]{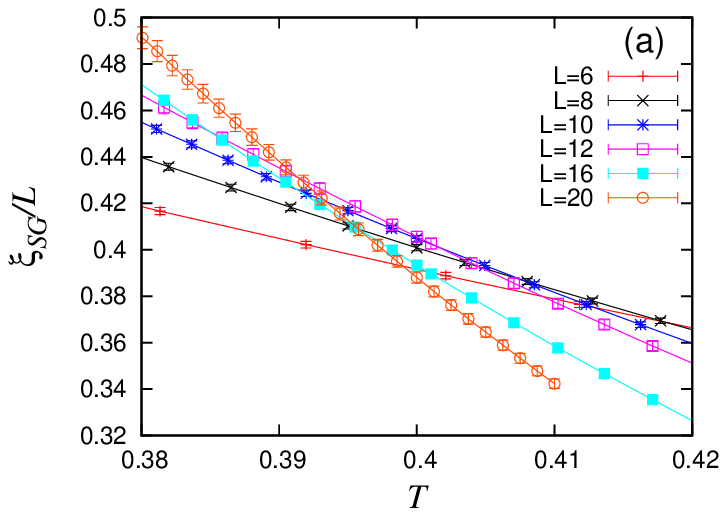}
\includegraphics[scale=1.1]{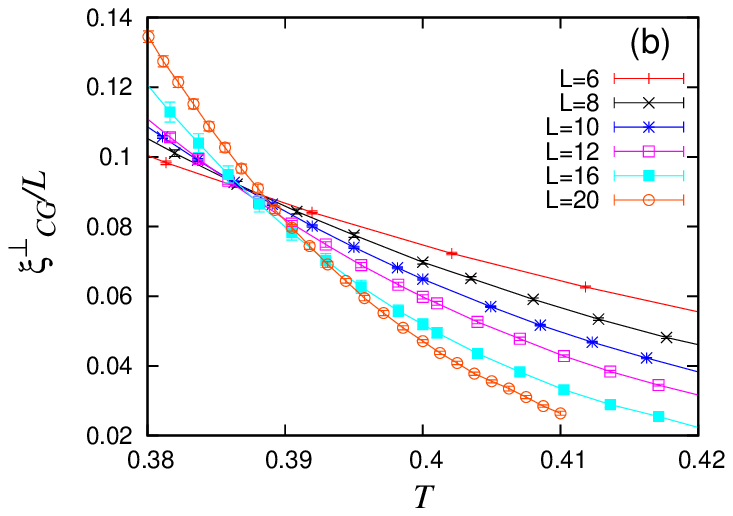}
\includegraphics[scale=1.1]{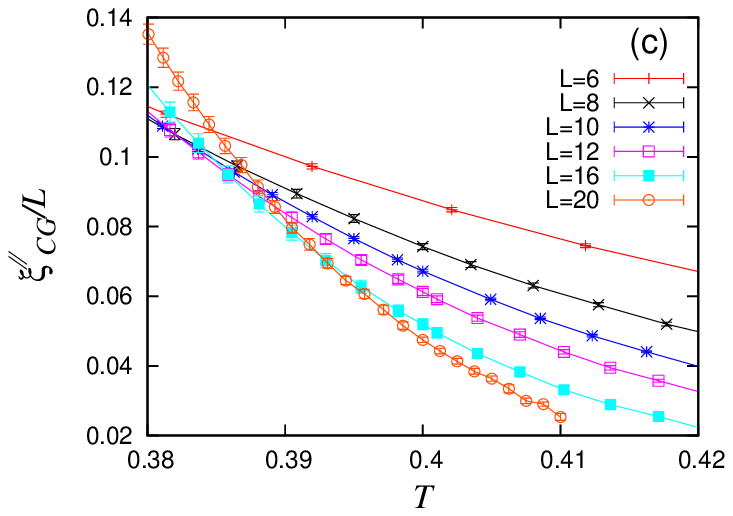}
\end{center}
\caption{
(Color online) Magnified views of the temperature and size dependence of the spin correlation-length ratio (a), of the perpendicular chiral correlation-length ratio (b), and of the parallel chiral correlation-length ratio (c). 
}
\end{figure}

 As an other indicator of the transition, we show in Fig.3 the Binder ratios for the spin (a) and for the chirality (b). The chiral Binder ratio $g_{CG}$ exhibits a negative dip. The data of different $L$ cross on the {\it negative\/} side of $g_{CG}$. A magnified view of $g_{CG}$ in the crossing-temperature region is shown in Fig.4.

 In contrast to $g_{CG}$, the spin Binder ratio $g_{SG}$ shown in Fig.3(a) exhibits no crossing in the investigated range of the temperature and the lattice size, monotonically decreasing with $L$. However, $g_{SG}$ develops a more and more singular shape with increasing $L$, a prominent peak appearing for larger $L$.

 In the $L\rightarrow \infty$ limit, the Binder ratios $g_{SG}$ and $g_{CG}$ should satisfy  here $g\rightarrow 0$ in the high-temperature phase, and $g=1$ at $T=0$. Hence, the asymptotic form of $g_{CG}$ in the $L\rightarrow \infty$ limit should be like the one as illustrated in the inset of Fig.3(b). In fact, such a form of $g$ is expected in a system with an ordered state exhibiting a continuous one-step-like replica-symmetry breaking (RSB) \cite{ImaKawa03b}.  A similar form of $g_{CG}$ was observed in 3D \cite{HukuKawa05,VietKawamura09a,VietKawamura09b}. Note that the one-step-like RSB discussed here is of a continuous type, in contrast to the one-step RSB of a discontinuous type often discussed in conjunction with structural glasses. In the latter case, the negative dip of $g_{CG}$ should exhibit a negative divergence at the transition, while such a negatively divergent behavior is not observed here.

\begin{figure}[ht]
\begin{center}
\includegraphics[scale=1.1]{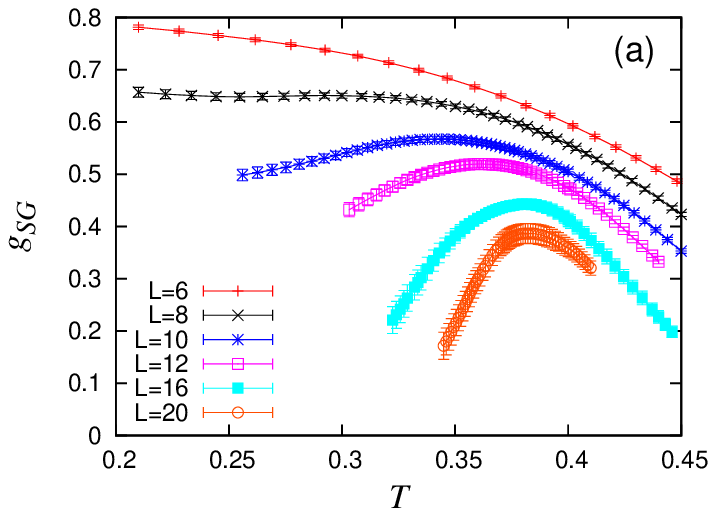}
\includegraphics[scale=1.1]{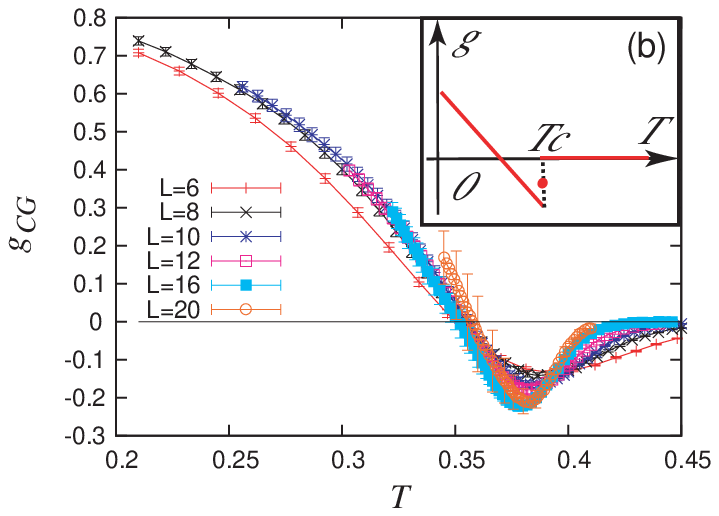}
\end{center}
\caption{
(Color online) The temperature and size dependence of the Binder ratio for the spin (a), and for the chirality (b). The inset of (b) is a behavior of $g_{CG}$ expected in the $L\rightarrow \infty$ limit.
}
\end{figure}
\begin{figure}[ht]
\begin{center}
\includegraphics[scale=1.2]{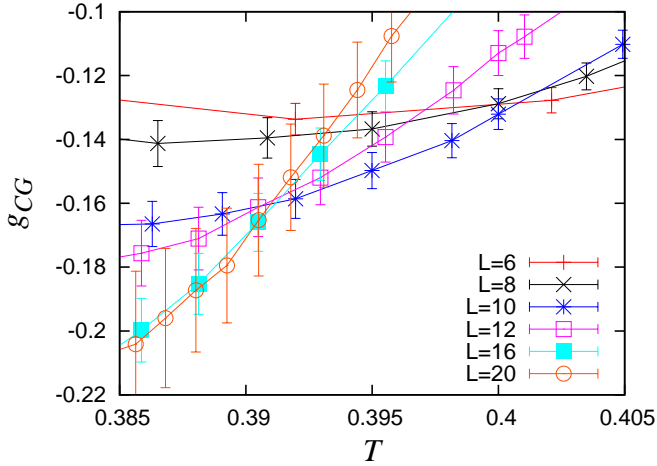}
\end{center}
\caption{
(Color online) A magnified view of the chiral Binder ratio $g_{CG}$ in the crossing-temperature range.
}
\end{figure}

 In order to estimate the bulk SG and CG transition temperatures quantitatively, we plot in Fig.5(a) the crossing temperatures $T_{cross} (L)$ of  $\xi_{SG}/L$ versus the inverse system size $1/L_{av}$ for pairs of the sizes $L$ and $sL$ with $s=2, 5/3$ and 5/4, where $L_{av}=\frac{1+s}{2}L$. Likewise, the crossing temperatures $T_{cross} (L)$ of $\xi_{CG}^\perp/L$, $\xi_{CG}^\parallel/L$ and $g_{CG}$ are plotted  versus $1/L_{av}$ in Fig.5(b).  The crossing temperature $T_{cross} (L)$ is expected to obey the scaling form,%
\begin{equation}
T_{cross}(L;s)=T_g + c_s L^{-\theta}, \ \ \  
\theta=\omega+\frac{1}{\nu} , 
\end{equation}
where $\nu$ is the correlation-length exponent and $\omega$ is the leading correction-to-scaling exponent. We fit our data of $T_{cross} (L;s)$ for the spin or for the chirality to the above form (11), to extract the transition temperature ($T_g=T_{CG}$ or $T_{SG}$) and the exponent $\theta$ ($\theta=\theta_{CG}$ or $\theta_{SG}$) for the spin or the chirality. 

\begin{figure}[ht]
\begin{center}
\includegraphics[scale=1.2]{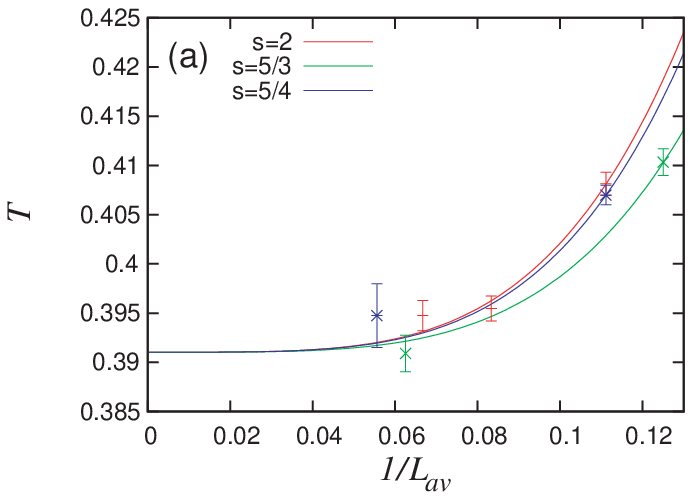}
\includegraphics[scale=1.2]{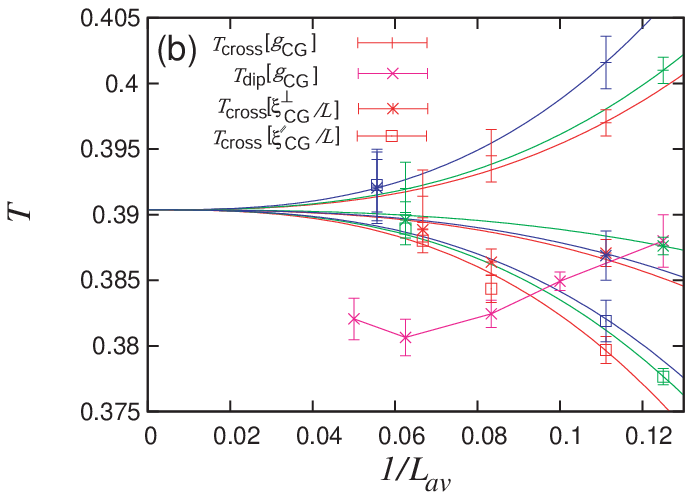}
\end{center}
\caption{
(Color online) The crossing temperatures $T_{cross}$ of several quantities between the sizes $L$ and $sL$ are plotted versus the inverse system size $1/L_{av}$ with $L_{av}=\frac{1+s}{2}L$: The spin correlation-length ratio  $\xi_{SG}/L$ in (a); the perpendicular chiral correlation-length ratio $\xi_{CG}^\perp/L$, the parallel chiral correlation-length ratio $\xi_{CG}^\parallel/L$, and the chiral Binder ratio $g_{CG}$ in (b). In (b), the dip temperatures $T_{dip}$ of the chiral Binder ratio $g_{CG}$ are also shown. Solid curves represent the fitting curves of the data on the basis of on Eq.(11) (see the text for details). The spin-glass and the chiral-glass transition temperatures are estimated to be $T_{SG}=0.391(2)$ and  $T_{CG}=0.390(1)$, respectively.
}
\end{figure}

 For the spin, we  perform a joint fit of $T_{cross}(L;s)$ of $\xi_{SG}/L$ for three different values of $s=2, \frac{5}{3}, \frac{5}{4}$, where the values of $T_{SG}$ and $\theta_{SG}$ are taken common while the values of $c_s$ be $s$-dependent. We then find an optimal fit for $T_{SG}=0.391(2)$ and $\theta_{SG}=4(2)$ with the associated $\chi^2$ value, $\chi^2$/DOF=0.73.

 For the chirality, we have several kinds of crossing temperatures $T_{cross}(L;s)$, {\it i.e.\/}, $T_{cross}(L;s)$ of $\xi_{CG}^\parallel/L$, $\xi_{CG}^\perp/L$ and $g_{CG}$. Then, we perform a joint fit of the data of $T_{cross}(L;s)$ of these three kinds of $T_{cross}(L;s)$, each with $s=2, \frac{5}{3}, \frac{5}{4}$, where $T_{CG}$ and $\theta_{CG}$ are taken common while the values of $c_s$  be $s$-dependent. We then get $T_{CG}=0.390(1)$ and $\theta_{CG}=2.4(4)$ with the associated $\chi^2$ value, $\chi^2$/DOF=0.51. 

 One sees from these results that the spin and the chiral transition temperatures agree within the error bars, {\it i,e,\/} $T_{SG}=T_{CG}$ within the accuracy of 1\%. This observation strongly suggests the absence of the spin-chirality decoupling in 4D, in contrast to the case of 3D where $T_{SG}$ lies below $T_{CG}$ by about $10\sim 15$\%. 

 For the CG transition, we have another indicator, {\it i.e.\/}, the negative-dip temperature $T_{dip}(L)$ of the chiral Binder ratio $g_{CG}$, which is expected to obey the scaling form,
\begin{equation}
T_{dip}(L;s)=T_g + c L^{-\frac{1}{\nu}}.
\end{equation}
 The data of  $T_{dip}(L)$ are also shown in Fig.5(b). As can be seen from the figure, $T_{dip}(L)$ changes its behavior with increasing $L$. It tends to {\it decrease\/} with $L$ for smaller sizes, while it tends to {\it increase\/} with $L$ for larger sizes of $L\gtrsim 16$. Indeed, such a non-monotonic size-dependence of $T_{dip}(L)$ is expected due to the following reason. For large enough $L$, the negative-dip temperature $T_{dip}(L)$ should lie {\it below\/} the crossing temperature  of $g_{CG}$, $T_{cross}(L)$. Since the exponents governing the asymptotic size dependence of $T_{dip}(L)$ and $T_{cross}(L)$ are $\theta$ and $1/\nu$ which satisfy the inequality $\theta > 1/\nu$ by definition, $T_{dip}(L)$ needs to approach $T_{CG}$ {\it from below\/} for large enough $L$. Hence, a bending-up behavior observed in $T_{dip}(L)$ for larger $L$ is a necessary changeover as expected from the argument above. 

 Anyway, this changeover in the observed size-dependence of $g_{dip}(L)$ makes a systematic extrapolation of  $T_{dip}(L)$ difficult. Nevertheless, as can be seen from Fig.5(b), our data of  $T_{dip}(L)$ for larger $L\geq 16$ seems fully consistent with the $T_{CG}$-value obtained above from the crossing temperatures. %In fact, the assumption $\nu_{CG}\simeq 1$ will be fully justified as we shall see below. 

 As mentioned above, the negative dip of $g_{CG}$ shown in Fig.3(b) is consistent with the occurrence of a one-step-like RSB \cite{HukuKawa05,VietKawamura09a,VietKawamura09b}. The corresponding spin Binder ratio $g_{SG}$ shown in Fig.3(b) also develops a more and more singular form with a peak structure appearing for larger $L$. If one recalls the fact that $g_{SG}$ takes a value unity at $T=0$ and approaches zero above $T_{SG}(=T_{CG})$ in the $L\rightarrow \infty$ limit, $g_{SG}$ is expected to develop a negative dip as in the case of $g_{CG}$. In the $L\rightarrow \infty$ limit, this negative dip temperature $T_{dip}$ should yield $T_{SG}$. Since $T_{SG}$ is likely to agree with $T_{CG}$ in 4D, a one-step-like RSB is expected to arise independently of the occurrence of the spin-chirality decoupling. In other words, in 4D, the Heisenberg SG is likely to exhibit a single SG transition without the spin-chirality decoupling. Yet, the SG (simultaneously CG) ordered state is peculiar in that the ordered state possesses a one-RSB-like nontrivial phase-space structure.

\section{V. Critical properties}

 In the previous section, we have demonstrated that, in 4D, the SG and the CG transitions are likely to take place simultaneously, {\it i.e.}, $T_{SG}=T_{CG}$. In this section, we study the critical properties of the transition on the basis of a finite-size scaling analysis of our MC data. In the absence of the spin-chirality decoupling, a natural expectation for the critical properties  is that, as usual, the spin is a primary order parameter of the transition. Then one should have $\nu_{SG}=\nu_{CG}$ and $\eta_{SG} < \eta_{CG}$. The latter corresponds to the fact that the spin is the primary order parameter and the chirality is the composite of the spin. 

 We first study the  critical properties of the spin by means of a finite-size scaling analysis of both $\xi_{SG}/L$ and $q^{(2)}_{SG}$. We employ the following finite-size scaling form with the leading correction-to-scaling term,

\begin{equation}
\frac {\xi_{SG}}{L}=\tilde X ((T-T_{SG})L^{1/\nu_{SG}})(1+aL^{-\omega_{SG}}),
\end{equation}
\begin{equation}
q_{SG}^{(2)}=L^{-(2+\eta_{SG})}\tilde Y((T-T_{SG})L^{1/\nu_{SG}})(1+a'L^{-\omega_{SG}}) ,
\end{equation}
where $a$ and $a'$ are numerical constants, while $\tilde X$ and $\tilde Y$ are appropriate scaling functions. The SG transition temperature $T_{SG}$ is fixed to $T_{SG}=0.39$ as determined in the previous section. 

 We begin with the finite-size scaling of $\xi_{SG}/L$ with $\nu_{SG}$ and $\omega_{SG}$ free fitting parameters. The best fit is obtained for $\nu_{SG}=1.0$ and $\omega_{SG}$=3.0. The resulting scaling plot is given in Fig.6(a). Inspecting the quality of the plot by eyes, we put the error bars as $\nu_{SG}=1.0(1)$ and $\omega_{SG}$=3(1). Note that these estimates of $\nu_{SG}$ and $\omega_{SG}$ are consistent with our above estimate of $\theta_{SG}=\omega_{SG}+\frac{1}{\nu_{SG}}=4(2)$. Next, with assuming  $\nu_{SG}=1$ and $\omega_{SG}=3$, we perform the finite-size scaling analysis of $q_{SG}^{(2)}$ to obtain $\eta_{SG}=-0.3(1)$. The resulting scaling plot is shown in Fig.6(b). 

 We also try the type of the extended finite-size scaling analysis proposed by Campbell {\it et al\/} where the scaling variables are chosen to take a matching  between the critical regime and the high temperature regime in order to get a wider scaling regime \cite{Campbell06}. The resulting exponent values turn out to be the same as those  obtained above by the standard analysis.

\begin{figure}[ht]
\begin{center}
\includegraphics[scale=1.1]{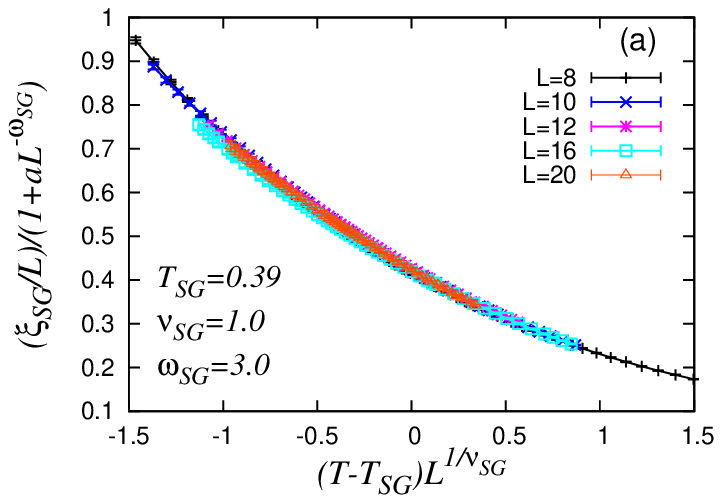}
\includegraphics[scale=1.1]{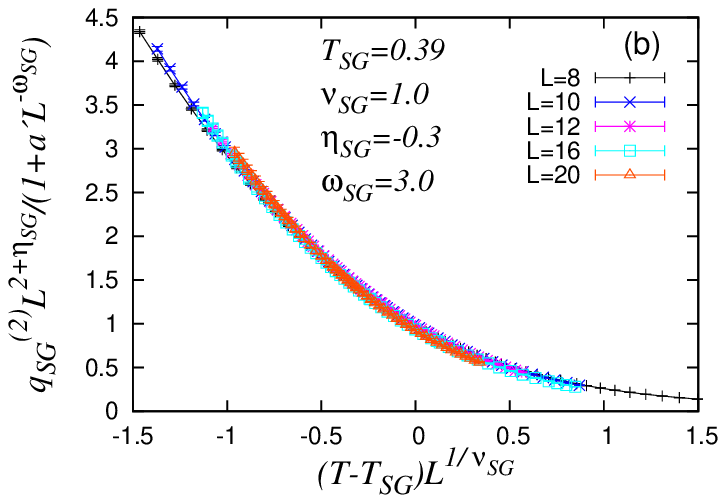}
\end{center}
\caption{
(Color online) Finite-size-scaling plots of the spin-glass correlation-length ratio $\xi_{SG}/L$ (a), and of the spin-glass order parameter $q_{SG}^{(2)}$ (b), where the leading correction-to-scaling effect is taken into account. The spin-glass transition temperature is fixed to $T_{SG}=0.39$ as determined in \S IV. The best fit for $\xi_{SG}/L$ is obtained with $\nu_{SG}=1.0$ and $\omega_{SG}=3.0$, while that for $q_{SG}^{(2)}$ is obtained with $\nu_{SG}=1.0$ (fixed) and $\eta_{SG}=-0.3$.
}
\end{figure}

 Similar scaling analysis is also applied to the chiral degrees of freedom to estimate the chiral correlation-length exponent $\nu_{CG}$ and the chiral anomalous-dimension exponent $\eta_{CG}$. The transition temperature is fixed to $T_{CG}=0.39$ as determined in the previous section. The finite-size scaling of the chiral correlation-length ratio yields $\nu_{CG}=1.0(1)$ and $\omega_{CG}=1.7(3)$. We get the same estimates even when we use either the perpendicular or the transverse CG correlation-length ratio. The resulting scaling plot for the perpendicular one is given in Fig.7(a). These estimates of $\nu_{CG}$ and $\omega_{CG}$ are consistent with our above estimate of $\theta_{SG}=\omega_{SG}+\frac{1}{\nu_{SG}}=2.3(4)$.  The finite-size scaling of the CG order parameter $q_{CG}^{(2)}$ with fixing $\nu_{CG}=1.0$ and $\omega_{CG}=1.7$ yields $\eta_{CG}=2.4(8)$. The resulting scaling plot is given in Fig.7(b).  We also try the extended finite-size scaling analysis a la Campbell {\it et al\/} \cite{Campbell06}. Again,  as in the case of the spin, the resulting exponent values turn out to be the same as those obtained above by the standard analysis.

\begin{figure}[ht]
\begin{center}
\includegraphics[scale=1.1]{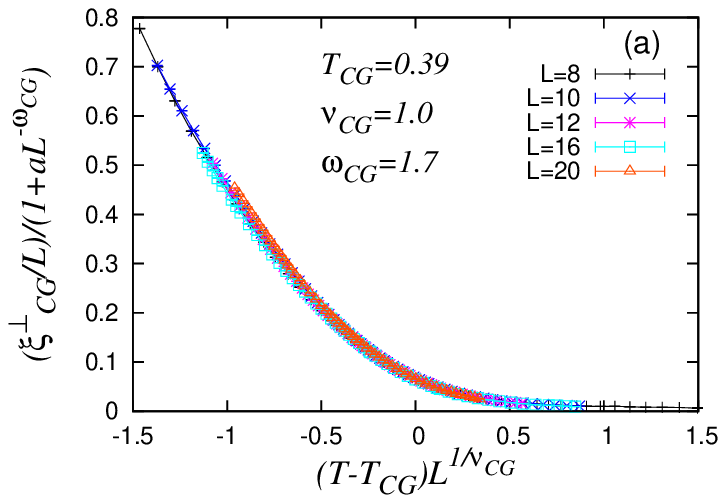}
\includegraphics[scale=1.1]{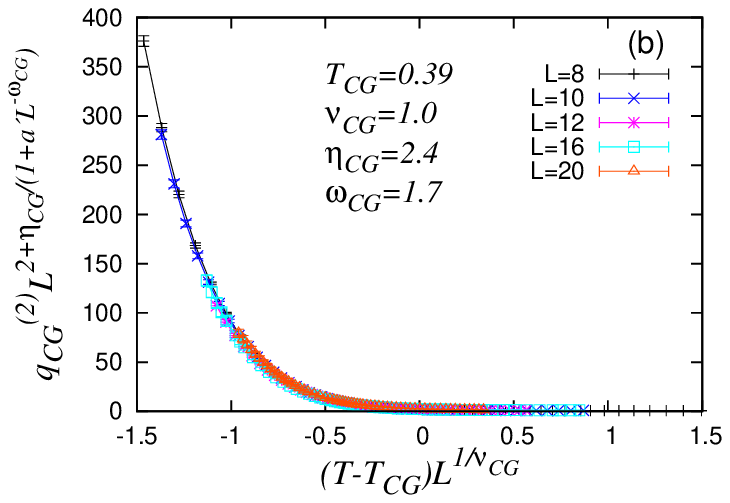}
\end{center}
\caption{
(Color online) Finite-size-scaling plots of the perpendicular chiral-glass correlation-length ratio $\xi_{CG}^\perp/L$ (a), and of the chiral-glass order parameter $q_{CG}^{(2)}$ (b), where the leading correction-to-scaling effect is taken into account. The chiral-glass transition temperature is fixed to $T_{CG}=0.39$ as determined in \S 4. The best fit for $\xi_{CG}^\perp /L$ is obtained with $\nu_{CG}=1.0$, while that for $q_{CG}^{(2)}$ is obtained with $\nu_{CG}=1.0$ (fixed) and $\eta_{CG}=2.4$. 
}
\end{figure}

 Combining the exponent estimates obtained above, we finally quote as our best estimates of the spin exponents,
\begin{equation}
\nu_{SG}=1.0\pm 0.1\ ,\ \ \ \eta_{SG}=-0.3\pm 0.1,
\end{equation}
while for the chirality exponents quote
\begin{equation}
\nu_{CG}=1.0\pm 0.1\ ,\ \ \ \eta_{CG}=2.4\pm 0.8.
\end{equation}
If one applies the standard scaling or hyperscaling relations, one can estimate other SG exponents as $\alpha\simeq -2.0$, $\beta_{SG} \simeq 0.85$, $\gamma_{SG}\simeq 2.3$, and $\delta_{SG}\simeq 3.7$, {\it etc\/}. 

 One sees from these estimates that the correlation-length exponents $\nu$ for the spin and for the chirality agree within the error bars, {\it i.e.\/}, $\nu_{SG}=\nu_{CG}$, which indicates the existence of only one diverging length scale at the transition. This observation is fully consistent with the absence of the spin-chirality decoupling in the 4D Heisenberg SG. Our data are also not incompatible with the relation $\omega_{SG} = \omega_{CG}$ within the error bars. By contrast, the anomalous-dimension exponents satisfy the inequality $\eta_{SG} < \eta_{CG}$, indicating that the spin is the primary order parameter as usual. If one applies the scaling relation to the CG exponents $\gamma_{CG}=(2-\eta_{CG})\nu_{CG}$, one would get the CG susceptibility exponent as $\gamma _{CG}=-0.4\pm 1.1$. The estimated value of $\gamma_{CG}$ means that the CG susceptibility does not diverge, or diverges only weakly, at the transition. This observation is again consistent with the view that the primary order parameter in 4D is the spin and the chirality is only composite.

 The obtained CG exponents values might be compared with the earlier estimates by Imagawa and Kawamura on the same model, {\it i.e.\/}, $\nu_{SG}=1.3(2)$ and $\eta_{SG}=-0.7(2)$ \cite{ImaKawa03a}. One sees that $\nu_{SG}$ agrees with our present estimate within the error bars, while $\eta_{SG}$ deviates somewhat. In view of the larger sizes employed in the present study as compared with those of ref.\cite{ImaKawa03a}, {\it i.e.\/}, $L\leq 20$ vs. $L\leq 10$, and also of larger number of independent samples, {\it e.g.\/}, 840 vs. 80 for $L=10$, our present estimate would be more trustable.

\section{VI. Summary and discussion}

 We studied equilibrium ordering properties of the 4D isotropic Heisenberg SG by means of an extensive MC simulation. By calculating various physical quantities including the correlation-length ratio, the Binder ratio and the glass order parameter up to the size as large as $L=20$ and down to temperatures well below $T_g$, we have found that $T_{SG}=0.391(2)$ is likely to coincide with $T_{CG}=0.390(1)$, which indicates that the spin and the chirality order simultaneously in the 4D Heisenberg SG, {\it i.e.\/}, the absence of the spin-chirality decoupling. If $T_{SG}$ and $T_{CG}$ are to differ, the distance in transition temperatures should be less than 1\%. We also studied the critical properties of the transition on the basis of the finite-size scaling analysis. The exponents were estimated to be $\nu_{SG}=1.0(1)$ and $\eta_{SG}=-0.3(1)$ for the spin, and  $\nu_{CG}=1.0(1)$ and $\eta_{CG}=2.4(8)$ for the chirality. Although the SG transition in 4D is usual in the sense that the spin is the primary order parameter, the standard exponent relations $\nu_{SG}=\nu_{CG}$ and $\eta_{SG} < \eta_{CG}$ being satisfied. Yet,  the SG transition is somewhat unusual in the sense that the low-temperature SG (simultaneously CG) ordered state exhibits a nontrivial phase-space structure, {\it i.e.\/}, a continuous one-step-like RSB. Note that the type of RSB is quite different from the one observed in the Ising SG, or the one observed in the mean-field limit of both the Ising and the Heisenberg SGs. 

 As mentioned in \S 1, possible correspondence between the orderings of the $d$-dimensional SR Heisenberg SG and of the 1D LR Heisenberg SG with a power-law interaction has been suggested in the literature. Although this correspondence is by no means exact, recent numerical studies both on the Ising and the Heisenberg SGs supported such correspondence. Indeed, Katzgraber {\it et al\/} proposed a formula for the $d$-$\sigma$ correspondence, a refined version of the one mentioned in \S 1 \cite{Katzgraber09},
\begin{equation}
d = \frac{2-\eta_{SG}}{2\sigma -1},
\end{equation}
where $\eta_{SG}$ is the spin anomalous-dimension exponent of the $d$-dimensional SR system.
% Note that Katzgraber et al originally dealed with the Ising SG which did not possess the chiral degree of freedom. 
Now, we have an estimate of $\eta_{SG}$ for the $d=4$ Heisenberg SG as $\eta_{SG}\simeq -0.3$. Substituting this into the r.h.s. of eq.(17) and putting $d=4$, we get $\sigma=0.79$. Together with the recent numerical estimate of the borderline value of $\sigma_c$ separating the spin-chirality coupling/decoupling regimes, $\sigma_c\simeq 0.8$ \cite{VietKawamura10a,VietKawamura10b}, the $d$-$\sigma$ correspondence suggests that the 4D lies very close to the borderline dimensionality of the spin-chirality coupling/decoupling, on the coupling regime only slightly. Such a view on the basis of the $d$-$\sigma$ correspondence seems fully consistent with our present MC results.

 In fact, the correspondence holds also for the critical exponents. In the $d$-$\sigma$ analogy, the exponent $\nu_{SG}$ of the 1D LR model should be related to that of the $d$-dimensional SR model via the relation, $\nu_{SG}$[1D-LR]=$d\times \nu_{SG}$[$d$D-SR] \cite{Larson}. Then, our 4D result suggests that the corresponding 1D LR model should be characterized by the exponent $\nu_{SG}\simeq 4\times 1.0=4$. Meanwhile, ref.\cite{VietKawamura10b} gave  $\nu_{SG}=3.6(5)$ and $\nu_{CG}=4.0(5)$  for $\sigma=0.8$ so that the expected relation is indeed satisfied.  All these results suggest that $d=4$ probably lies fairly close to the borderline dimensionality of the spin-chirality decoupling/coupling, may even lie just at the border.

\begin{acknowledgments}
The authors are thankful to T. Okubo and T. Obuchi for useful discussion. This study was supported by Grant-in-Aid for Scientific Research on Priority Areas ``Novel States of Matter Induced by Frustration'' (19052006 \& 19052007). We thank ISSP, University of Tokyo, YITP, Kyoto University, and Cyber Media Center, Osaka University for providing us with the CPU time.
\end{acknowledgments}

\end{document}